# Translational and reorientational dynamics in carboxylic acid-based deep eutectic solvents

A. Schulz[1,a], K. Moch[2], Y. Hinz[2], P. Lunkenheimer[1], and R. Böhmer[2]

**AFFILIATIONS**

[1]Experimental Physics V, Center for Electronic Correlations and Magnetism, University of Augsburg, 86159 Augsburg, Germany

[2]Fakultät Physik, Technische Universität Dortmund, 44221 Dortmund, Germany

[a]**Author to whom correspondence should be addressed:** arthur.schulz@physik.uni-augsburg.de

**ABSTRACT**

The glass formation and the dipolar reorientational motions in deep eutectic solvents (DESs) are frequently overlooked, despite their crucial role in defining the room-temperature physiochemical properties. To understand the effects of these dynamics on the ionic conductivity and their relation to the mechanical properties of the DES, we conducted broadband dielectric and rheological spectroscopy over a wide temperature range on three well-established carboxylic-acid-based natural DESs: oxaline, maline, and the eutectic mixture of choline chloride with phenylacetic acid (phenylaceline). In all three DESs, we observe signs of glassy freezing in the temperature dependence of their dipolar reorientational and structural dynamics, as well as varying degrees of motional decoupling between the different observed dynamics: Maline and oxaline display a breaking of the Walden rule near the glass-transition temperature, while the relation between the dc conductivity and dipolar relaxation time in both maline and phenylaceline is best described by a power law. The glass-forming properties of the investigated systems not only govern the orientational dipolar motions and rheological properties, which are of interest from a fundamental point of view, but they also affect the dc conductivity, even at room temperature, which is of high technical relevance.

## I. INTRODUCTION

Since the pioneering work by Abbott *et. al.*,[1] the material class of the deep eutectic solvents (DESs) continues to receive a lot of attention due to their versatile physiochemical properties and possible applications.[2,3,4,5,6,7,8] These eutectic mixtures of Lewis and Brønsted acids and bases, or hydrogen-bond donors (HBDs) and acceptors (HBAs), are formed in a distinct molar composition range, possess a eutectic point temperature below that of an ideal liquid mixture, and appear mostly liquid around room temperature.[9,10,11] The melting point depression, with respect to each of their components,[12] and the formation of DESs, is intimately related to the development of hydrogen bonds.[13,14]

DESs are categorized based on their complexing agents[3,9] and exhibit an immense pool of possible constituents and mixing options, rendering them highly versatile in regards to their physical properties.[15,16,17,18] The density, viscosity, ionic conductivity, surface tension, and polarity can all be tailored for specific applications by suited source materials, their mixing ratio, and the water-content in the compound.[3,19,20,21] This tunability in combination with the availability of cheap components make DESs vastly economic and is also the reason why they are often compared to the well-established ionic liquids (ILs). The latter also feature a wide composition range in the liquid state as well as low volatility.[11,16,22,23,24] Concerning sustainability, however, most DESs are superior,[23,25] particularly in the case of the so-called natural DESs (NADES). This important subgroup of DESs consists of eutectic mixtures formed by natural primary metabolites, such as organic acids, sugars, amino acids, and choline derivates.[3,26] Apart from the superior properties of conventional DESs, NADESs exhibit excellent recyclability, biodegradability, low toxicity, and ease of preparation from low-cost materials.[27,28,29] With these qualities, NADESs are fully in line with the goals of green chemistry[30] and surpass conventional organic solvents which are often flammable, volatile, and toxic.[31,32,33] This remarkable range of valuable and tunable physiochemical properties makes NADESs suitable for a variety of applications, ranging from effective extractants and utilization in the pharmaceutical industry as well as for food processing,[21,26,34] to electrochemical applications.[35,36,37,38,39]

Facing current energy-storage challenges,[40,41,42] the use of DESs as electrolytes in batteries is of special interest. Their environmental, economic, and safety advantages renders them promising alternatives to presently used materials.[36,43] Besides the ability to reversibly store and remove ions in metallic electrodes,[44,45] the direct current (dc) conductivity at room temperature represents a major benchmark for electrolyte properties. Indeed, in this respect several DESs reach technical relevance, which implies $\sigma_{dc} > 10^{-4}~\Omega^{-1}~cm^{-1}$.[38,46,47] However, to challenge energy densities currently reached in modern mixed organic carbonate electrolytes, it is necessary to better understand the ionic charge-transport in DES materials.[48,49]

In the present work, we focus on the ionic conductivity in relation to the dipolar reorientation processes and the mechanical properties of three NADESs formed by a carboxylic acid and the quaternary salt choline chloride (ChCl). In Abbott's study[1], maline (malonic acid + ChCl, 1:1 molar ratio), oxaline (oxalic acid + ChCl, 1:1 molar ratio), and the DES formed between phenylacetic acid and ChCl at the eutectic point (2:1 molar ratio; termed phenylaceline in the following) are found to have the highest conductivity around room temperature, in descending order, among nine investigated DESs. Oxaline is of special interest as it also exhibits the highest conductivity for a given viscosity, in marked contrast to phenylaceline which features the lowest conductivity in that regard (together with the phenylpropionic acid-based DES).[1] Carboxylic acids exhibit larger electric dipole moments compared to other organic compounds, allowing for a stronger hydrogen bonding[50] which explains their ease of complex formation with ChCl. The different molecular shapes and dipole moments of the chosen carboxylic acids[50,51,52] promise interesting insights into the interplay of their ionic and molecular dynamics. We utilize dielectric spectroscopy to simultaneously detect translational ionic and rotational dipolar motions, shear rheological spectroscopy to analyze structural dynamics, as well as differential scanning calorimetry (DSC) to access the glass-transition temperature. As previously shown for other DESs,[53,54,55,56,57] this combined approach is useful to clarify the relation between ionic, reorientational, and structural dynamics, as well as the role played by glass formation.

## II. EXPERIMENTAL DETAILS

The NADESs oxaline, maline, and phenylaceline were prepared by mixing choline chloride with the respective carboxylic acids (oxalic, malonic, and phenylacetic acid) using a magnetic stirrer. The molar compositions at the respective eutectic points (1:1, 1:1, and 1:2, respectively) are chosen based on the phase diagrams published by Abbott *et. al.*.[1] The given compositions maximize the freezing point depression relative to the initial constituents, causing maline and phenylaceline to even appear liquid around room temperature. Prior to the synthesis, choline chloride was dried under vacuum at 40°C for 48 h to ensure no moisture was present. In addition, all components were mortared before mixing to facilitate the formation of the eutectics. To prevent the previously reported formation of esters, HCl, and water in the NADESs at elevated temperature,[24,58,59] heating was avoided during their preparation, where possible. Only for oxaline, the synthesis required continuous stirring and heating at 75°C for 4 h. All other solvents were mixed at room temperature using a magnetic stirrer until the mixtures were clear, liquid, and without any visually detectable residuals. To further avoid moisture uptake, all samples were kept in dry nitrogen during synthesis, storage, and all performed experiments. All measurements, but the ones explicitly specified, were conducted directly after the synthesis, to avoid the formation of impurities.[24] Water in a DES influences the structural, dielectric, and rheologic properties of the solvent, as has been frequently investigated.[19,21,53,60,61] This finding is why the water content of each sample was determined by coulometric Karl-Fischer titration. The latter leads to water contents of significantly less than 0.1 wt% for maline and phenylaceline after synthesis. For oxaline, higher water contents of the order of $1-2$ wt% were detected. For the rheological measurements, all DESs were prepared immediately prior to the experiments and in the same manner as for the dielectric measurements. In the maline and phenylaceline samples the concentration of water is negligible. In oxaline, however, water, and thus further impurities arising from the esterification reaction, do exist in notable quantities, which is assessed in more detail in the data evaluation, below.

The dielectric measurements were conducted applying two methods: A frequency response analyzer (Novocontrol Alpha-analyzer) was used in the range between 0.1 Hz and about 10 MHz and an impedance analyzer (Keysight Technologies E4991B) featuring a coaxial reflectometric setup[62] in the frequency window between 1 MHz and 3 GHz. The dielectric spectra of the real part of the permittivity, obtained from the first method, are rescaled by subtracting a constant value arising from stray capacitance to agree with the spectra obtained from the more accurate high-frequency measurements. The temperature in the cryostat (Novocontrol Quatro) holding the sample cell was controlled by a continuous, heated $N_2$-gas flow from a dewar, allowing for a slow temperature variation of 0.25 K/min and the acquisition of dielectric data in a broad temperature range. A simultaneous, but independent measurement of the real part of the capacitance $C'$ and the conductance $G'$ yields the geometry-independent and frequency-dependent complex dielectric permittivity $\varepsilon^*(\nu) = \varepsilon'(\nu) - \varepsilon''(\nu)$ and the real part of the electrical conductivity $\sigma'(\nu)$. To address sample degradation over time caused by esterification reactions,[59] we conducted additional dielectric and rheological spectroscopy measurements on maline 7 and 28 days after the initial sample preparation. During the intervals between these experiments, the sample was kept in a dry $N_2$ atmosphere at ambient temperatures. However, we did not observe any significant changes in $\sigma_{dc}$, $\langle \tau_\varepsilon \rangle$, or $\langle \tau_J \rangle$ (supplementary material, Fig. S1) and thus conclude that chemical degradation is not relevant within the studied time frame.

The shear rheological data were measured using an MCR 502 rheometer in combination with an EVU20 evaporation unit and a CTD 450 oven from Anton-Paar. All measurements employed a 4 mm diameter parallel plate geometry (PP04). The shear oscillation frequencies covered a range from 7 mHz to 70 Hz with shear amplitudes adjusted such that the mechanical response was in the linear-response regime. The temperature was stabilized within 0.2 K for each measurement. While filling the plate geometry, a brief contact between sample and ambient air cannot completely be avoided. We minimized this contact time by loading the sample in less than 30 s. For each sample, spectra were remeasured at the highest temperatures after all spectra at lower temperatures were taken. In particular, for maline and oxaline a difference to the initial measurement did not appear, while for phenylaceline the two sets of data deviated significantly. Subsequent measurements confirmed that the behavior of the phenylaceline sample is not stationary. Therefore, only the rheological data for maline and oxaline were used to analyze their frequency-dependent spectra.

Furthermore, differential scanning calorimetry (DSC) measurements were performed using a DSC8500 from PerkinElmer. With a scanning rate of 10 K/min, the glass-transition temperature $T_g^{DSC}$ could be determined by identifying the onset of the characteristic, step-like increase in the heating data of each sample.

## III. RESULTS AND DISCUSSION

### A. Dielectric spectroscopy

Figure 1 presents the spectra of the real ($\varepsilon'$) and imaginary ($\varepsilon''$) part of the dielectric permittivity, along with the real part of the conductivity $\sigma'$, obtained for maline. These spectra are discussed as examples, as qualitatively similar interpretations and results were observed for oxaline and phenylaceline, which are provided in the supplementary material (Figs. S2 and S3).

The most prominent feature of the dielectric-constant spectra [Fig. 1(a)] is the considerable increase to "colossal" values ($\gtrsim 10^3$)[63] at low frequencies and high temperatures which is accompanied by the indication of a peak in $\varepsilon''(\nu)$ (b). This non-intrinsic behavior is associated with so-called Maxwell-Wagner (MW) relaxations and is frequently observed in dielectric investigations of materials revealing significant charge transport.[63,64,65,66] In the case of ionic conductors, MW relaxations can arise from the accumulation of ions at the electrodes resulting in the formation of thin insulating layers with huge capacitances.[64,65,67] Such electrode blocking leads to a reduced conductivity at elevated temperatures, as evidenced by the decrease towards lower frequencies that is observed in the spectra of the real part of the conductivity [Fig. 1(c)]. At the highest measured temperatures, $\varepsilon'(\nu)$ exhibits an additional increase [e.g., below about 10 Hz for the 269 K curve in Fig. 1(a)] which is accompanied by similar effects in the dielectric loss and real part of the conductivity, as previously mentioned for the primary MW relaxation. This observation may be explained by the presence of a second, slower ion-transport mechanism within the electrode blocking layer, which can be attributed to a low, remaining ionic mobility in that region.[65,67]

At higher frequencies and/or lower temperatures, the dielectric constant exhibits an additional, smaller step-like decrease towards increasing frequencies, well observable in the magnified view provided by the inset of Fig. 1. This spectral contribution can be attributed to the reorientational motion of molecular dipoles, commonly termed $\alpha$ relaxation. Considering that 50 mol% of maline consists of dipolar molecules, the presence of a dipolar relaxation feature is expected, as reported in various other DESs containing dipolar constituents.[53,54,55,56,57,68,69,70] Upon cooling, the rotational dynamics of the dipoles gradually slows down, leading to the observed shift of the $\alpha$ relaxation step towards lower frequencies [Fig. 1(a)]. Typically, a relaxation step in $\varepsilon'(\nu)$ is accompanied by a peak in $\varepsilon''(\nu)$. However, in the dielectric loss spectra of maline, as depicted in Fig. 1(b), only a change of slope is visible (e.g., at $\nu \approx 10^4$ Hz for the 223 K curve), which is attributed to the high dc conductivity of the sample that superimposes the relaxation peak and contributes to the spectrum via $\varepsilon''_{dc} = \sigma_{dc}/(2\pi\nu\varepsilon_0)$. This phenomenon has been reported previously for both ILs[71,72,73] and highly conducting DESs.[53,54,55,56,70]

At even lower temperatures, $T < 205$ K, the spectral features of the primary relaxation shift out of the accessible frequency window, revealing two minor, secondary processes, referred to as $\beta$ and $\gamma$ relaxations. Both processes emerge at higher frequencies compared to the primary relaxation, have only weak amplitudes, and are best visible in the dielectric loss spectra [Fig. 1(b)]. The $\beta$ relaxation manifests as a gentle shoulder at the right flank of the $\alpha$ relaxation (e.g., at about 100 Hz for the 199 K curve). The broad $\gamma$ peak is only revealed at the lowest temperatures (e.g., at 300 Hz for the 147 K curve). Such secondary, intrinsic contributions to the spectra occur quite universally in dipolar glass-forming materials.[74,75] Their detailed treatment, however, is beyond the scope of the present work.

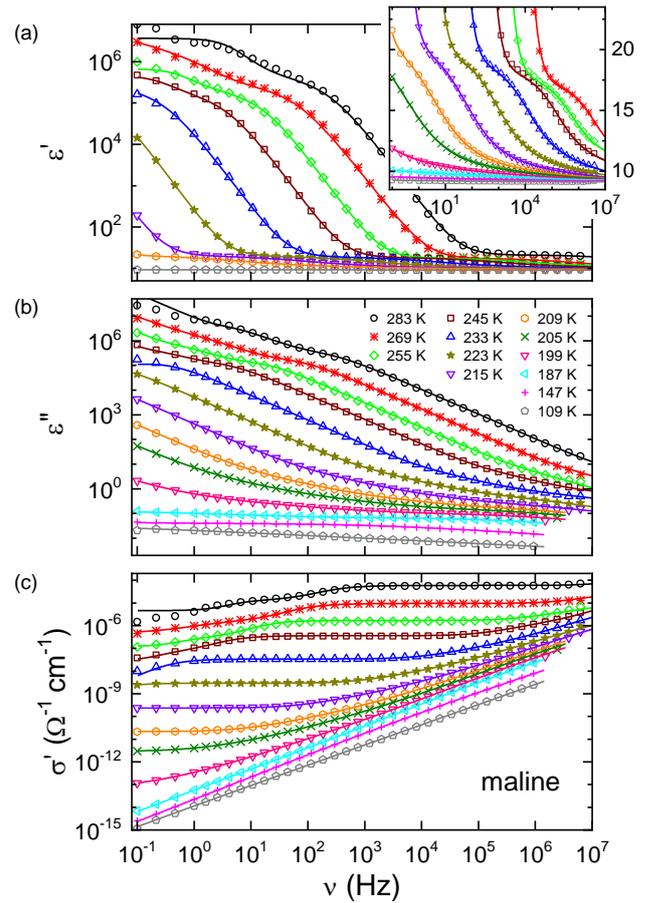

FIG. 1. Frequency dependence of the dielectric constant $\varepsilon'$ (a), dielectric loss $\varepsilon''$ (b), and the real part of conductivity $\sigma'$ (c), measured at various temperatures for maline. To enhance the readability, some temperatures have been omitted, especially in (a). The inset provides a magnified view of the dipolar relaxation steps in $\varepsilon'(\nu)$. The solid lines represent fits using an equivalent circuit approach, incorporating a distributed RC circuit to account for electrode blocking, intrinsic primary and secondary relaxation processes, as well as a contribution from dc conductivity as described in the text. The fitting procedure involved simultaneous fitting of the real and imaginary parts of the permittivity, with the shown conductivity fits calculated using $\sigma' = \varepsilon''2\pi\nu\varepsilon_0$.

Lastly, at frequencies above the decline associated with the MW relaxations, the $\sigma'$ spectra [Fig. 1(c)] exhibit plateaus that reflect the ionic dc conductivity $\sigma_{dc}$. Upon cooling, they strongly decrease by several orders of magnitude, indicating the pronounced temperature dependence of the charge-transport mechanism. At higher frequencies a steep increase in $\sigma'$ becomes apparent. It is dominated by the intrinsic relaxations discussed above, which contribute to $\sigma'$ due to the close relation of dielectric loss and conductivity, expressed by $\sigma' \sim \varepsilon''\nu$. It should be noted that the rise of $\sigma'$ at high frequencies [Fig. 1(c)] may also be attributed to ac conductivity caused by ionic hopping. Several models that deal with the ion charge transport, such as Dyre's random free-energy barrier model (RBM)[76,77] or Funke's jump relaxation model,[78] predict such a behavior. However, these models do not account for the reorientational dynamics of the dipolar constituents of the DESs, which represent a significant fraction of the components. Although an ac-conductivity contribution to the spectra, superimposed by the dominating dipolar relaxation response, cannot be fully excluded, we are able to account for all spectral features by fitting the data with contributions from dc conductivity and dipolar relaxations only. For further elaboration on these topic and examples of various fits of DES spectra, which include the RBM's predicted contributions, we refer the reader to previous publications.[54,55,56]

To conduct a quantitative analysis of the obtained spectra, the dielectric constant and loss are simultaneously fitted using an equivalent-circuit model, as outlined in Ref. 65. Following Occam's razor, the fitting procedure utilizes only the relevant contributions to the spectra within the observed frequency range at each temperature. This approach minimizes the number of parameters required for the fits and reduces the errors of the quantities obtained this way. The resulting fits are presented as lines in Fig. 1 as well as in Figs. S2 and S3 (supplementary material) showing the spectra for oxaline and phenylaceline. To model the contributions of the electrode polarization, up to two distributed RC circuits connected in series to the bulk are employed, as described in Ref. 65. The empirical Cole-Cole (CC) function [Eq. (1) with $\beta = 1$], resulting in symmetric loss peaks,[79] effectively describes the $\alpha$ relaxations of maline and phenylaceline.

$$\varepsilon^* = \varepsilon_\infty + \frac{\sigma_{dc}}{i\varepsilon_0 2\pi\nu} + \frac{\Delta\varepsilon}{[1 + (i2\pi\nu\tau)^{1-\alpha}]^\beta} \quad (1)$$

Equation (1) represents the Havriliak-Negami equation[80] where $\varepsilon_\infty$ is the high-frequency value of the dielectric constant, $\varepsilon_0$ the permittivity of free space, and $\Delta\varepsilon$ the relaxation strength. $\alpha$ and $\beta$ are width parameters that determine the symmetric and asymmetric broadening of the loss peak, respectively. For oxaline, the Cole-Davidson (CD) function[81] [Eq. (1) with $\alpha = 1$] is used, but a CC function provides an equally good description of the spectra (see Fig. S3 in the supplementary material). Typically, the CC function is reserved for secondary relaxations, whereas it is only rarely applied to the $\alpha$ relaxation, which usually exhibits asymmetric peaks. Nonetheless, in the DES glyceline, the $\alpha$ relaxation was also found to be accurately described by a CC function,[55] just as in the present scenario.

## B. Shear rheology

The rheological observable which can best be compared to the dielectric permittivity $\varepsilon^*$ is the complex shear compliance $J^* = J' - iJ''$. The correspondence of these two quantities has turned out useful for a range of different ion conducting glassformers[55,57,82,83,84] and therefore will also be exploited here. In the linear response regime, $J^*$ is the inverse of the complex shear modulus, $G^* = G' + iG'' = 1/J^*$. In Fig. 2 we show the storage compliance $J'$, the loss compliance $J''$,
and the loss modulus $G''$ of maline for temperatures between 194 and 211 K. The analogous plots for oxaline are presented in the supplementary material (Fig. S4).

The dielectric permittivity $\varepsilon^*$ and the rheological compliance $J^*$ display similar features. The storage compliance has plateaus at high and at low frequencies with values $J_\infty$ and $J_s = J'(0)$, respectively, where the static compliance $J_s$ is also called recoverable compliance. The plateaus in $J'$ are separated by a step of size $\Delta J = J_s - J_\infty$. One recognizes that the retardation times $\tau_J$, read out at the inverse of the frequencies at which the inflection points in $J'(\nu)$ appear, are longer than the relaxation times $\tau_G = 1/(2\pi\nu_{peak})$ which are assessed from the peaks in $G''$. This difference of modulus and susceptibility (= compliance) timescales is well known.[82] In Fig. 2(b) relaxational features cannot directly be discerned from the loss compliance $J''$. Merely, the fluidity contribution dominates the spectra and masks any loss peaks. Similarly, the loss permittivity $\varepsilon''$ is often superimposed by sufficiently strong conductivity contributions so that sometimes a dielectric loss peak is not visible.

We find that the shear data in Fig. 2 can be fitted with a CD function[81] augmented by a viscosity term

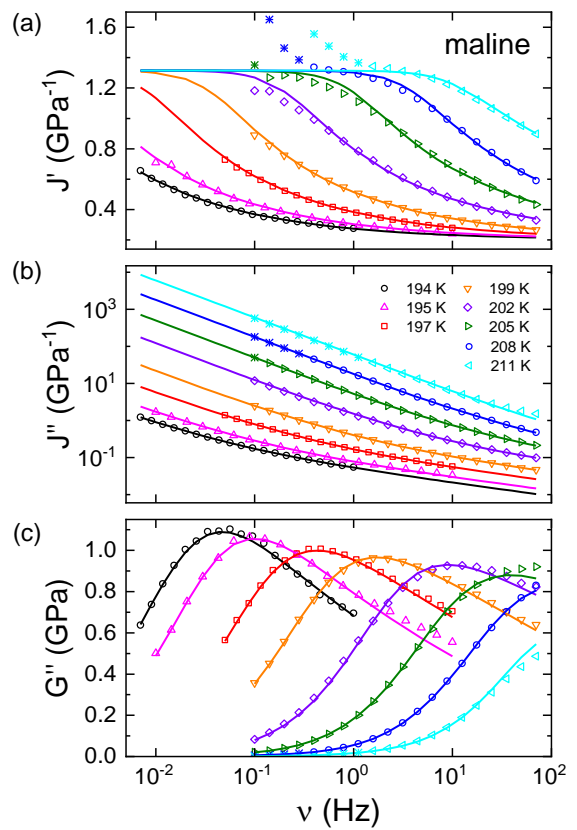

**FIG. 2.** Rheological spectra of maline in representations of (a) the storage compliance $J'$, (b) the loss compliance $J''$, and (c) the loss modulus $G''$. The solid lines in (a) and (b) result from simultaneous fits of $J'$ and $J''$ on the basis of Eq. (2) with fixed parameters $J_\infty = 0.2 \pm 0.02$ GPa$^{-1}$, $\Delta J = 1.1 \pm 0.1$ GPa$^{-1}$, and $\beta_J = 0.34 \pm 0.02$. for all temperatures. Only the data presented by open symbols are used for the fits while data not considered for the fits are shown as asterisks. This concerns data which are affected by the low-torque resolution limit of the rheometer. The solid lines in (c) are calculated from the fit curves via $G'' = J''/(J'^2 + J''^2)$.

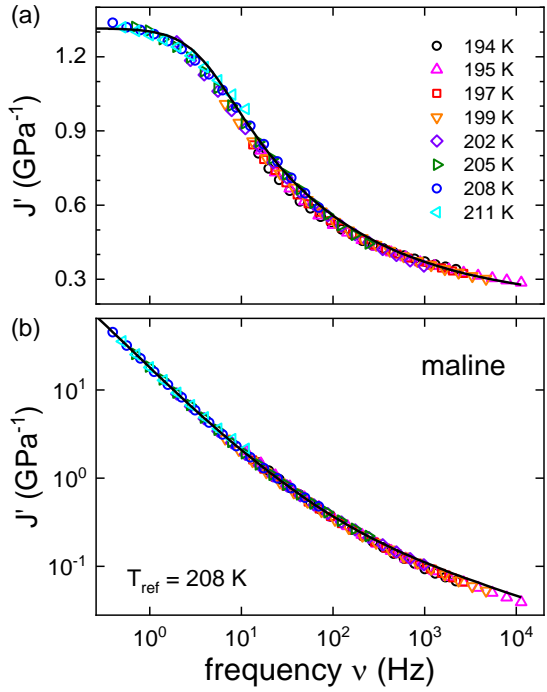

**FIG. 3.** (a) Storage and (b) loss compliance master curves for maline, constructed on the basis of the data shown in Fig. 2. The reference temperature is $T_{\text{ref}} = 208$ K. The solid lines are calculated using Eq. (2).

$$J^*(\nu) = J_\infty + \frac{\Delta J}{(1 + i2\pi\nu\tau_J)^{\beta_J}} + i\frac{1}{2\pi\nu\eta_0} \quad . \quad (2)$$

From $\tau_J$ and the width parameter $\beta_J$, the average retardation time $\langle\tau_J\rangle = \tau_J\beta_J$ can be extracted. Furthermore, $\eta_0$ denotes the steady-state viscosity.

In a first step, the data shown in Fig. 2 were fitted with free parameters, although complete $J'$ relaxation steps are seen in the rheological data for only a few temperatures. Therefore, it cannot be expected that this procedure yields reliable parameters for all temperatures. Consequently, in a second step, we fixed $J_\infty$, $\Delta J$, and $\beta_J$ within about 10% of their mean values. As Fig. 2 shows, the corresponding fits describe the data very well. Since the dielectric spectra of maline could be fitted in terms of a CC function, we also tested this approach for the rheological data of this DES. Here, however, the CD function consistently provides a superior description. For oxaline, the rheological data and the fits are shown as supplementary material (Fig. S4).

In Fig. 3 we present compliance master curves generated on the basis of the maline data shown Fig. 2. The master curves were obtained by first shifting the loss factors $\tan\delta = G''/G'$ onto each other along the frequency direction, with the spectrum recorded for $T = 208$ K as reference.[85] After applying the frequency shifts to the compliance spectra, minor vertical adjustments (±10%) were performed. Since this procedure generates consistent compliance master curves, the assumption that $J_\infty$, $\Delta J$, and $\beta_J$ are approximately independent of temperature is well justified.

## C. Coupling of reorientational and translational dynamics

Based on the fits of the dielectric spectra, Fig. 4 displays Arrhenius plots for the temperature-dependent dc conductivity $\sigma_{\text{dc}}$ (a) and the average relaxation time $\langle\tau_\varepsilon\rangle$ (b) of all three DESs studied. Notably, for all of them both quantities exhibit significant departures from a simple, thermally activated Arrhenius behavior. Such super-Arrhenius behavior is a distinctive characteristic of glass-forming liquids[75,86] and has been observed in various other DESs.[53,54,55,56,57,68,70] In order to characterize this non-Arrhenius freezing of the molecular and ionic motions upon cooling, Angell's version of the empirical Vogel-Fulcher-Tammann (VFT) law[87,88,89] is utilized,

$$q = q_0\exp\left(\frac{D_q T_{\text{VF}}}{T - T_{\text{VF}}}\right) \quad . \quad (3)$$

Here, $q$ represents $\sigma_{\text{dc}}$, $\langle\tau_\varepsilon\rangle$, or $\eta$, $q_0$ is the pre-exponential factor, $D_q$ is the strength parameter determining the degree of deviation from Arrhenius behavior,[90] and $T_{\text{VF}}$ is the divergence or Vogel-Fulcher temperature. The solid lines in Fig. 4 fit the experimental data well, with the parameters listed in Table 1. From the DSC measurements, the glass-transition temperatures $T_g^{\text{DSC}}$ of oxaline, maline, and phenylaceline were determined to be 224, 199, and 208 K, respectively. Making the common assumption $\langle\tau\rangle(T_g) \approx 100$ s, they are included in Fig. 4(b). The extrapolated VFT fits are consistent with $T_g^{\text{DSC}}$ for oxaline and maline, but significant deviations occur for phenylaceline. However, an alternative VFT fit that includes the value at $T_g^{\text{DSC}}$ (dashed line in Fig. 4(b)), demonstrates that the latter is compatible with the dielectric data of phenylaceline, as well.

Comparing the average dipolar relaxation times among the three DESs in Fig. 4(b) shows that they exhibit varying degrees of deviations from Arrhenius behavior. The fragility parameter $m_\tau$,[91] which can be derived from the strength parameter $D_\tau$ via $m_\tau = 16 + 590/D_\tau$,[92] reflects this trend (Table 1). The differences between the systems are most evident in the Angell plot[90] shown in the inset of Fig. 4, where it becomes clear that maline deviates strongest from an Arrhenius behavior. With a fragility parameter of 107, maline is categorized as a fragile glass-forming liquid. In contrast, oxaline and phenylaceline exhibit fragility values of 33 and 47, respectively, indicating a milder departure from a thermally activated behavior. In general, the fragility of a liquid can be interpreted as a measure of the complexity of its energy landscape.[93] Carboxylic-acid based DESs are known to form complex H-bonded molecule-ion structures. This was reported for maline[94,95] and for mixtures of choline chloride and oxalic acid.[96,97] Differences in the formation of such supramolecular structures for the three investigated DESs most likely play a role for their differences in fragility. Moreover, the number of different conformers in the HBDs could be a key factor behind the different fragilities. For a higher number of conformers, the possible local configurations in phase space also increases, resulting in a more complex energy landscape. Our experimental findings are in harmony with this line of thought, as different numbers of conformers are observed among the carboxylic acids investigated, with malonic acid having the most of them.[98,99,100]

| | $q$ | $T_{\text{VF}}$(K) | $D_q$ | $q_0$ |
|---|---|---|---|---|
| oxaline | $\langle\tau_\varepsilon\rangle$ | 121 | 34.6 | $1.7 \times 10^{-16}$ s |
| | $\rho_{\text{dc}}$ | 155 | 16.4 | $1.9 \times 10^{-3}$ $\Omega^{-1}$ cm$^{-1}$ |
| | $\eta$ | 159 | 15.3 | $3.9 \times 10^{-7}$ Pa s |
| maline | $\langle\tau_\varepsilon\rangle$ | 164 | 6.5 | $2.0 \times 10^{-11}$ s |
| | $\rho_{\text{dc}}$ | 152 | 9.7 | $3.1 \times 10^{-1}$ $\Omega^{-1}$ cm$^{-1}$ |
| | $\eta$ | 147 | 10.7 | $3.7 \times 10^{-7}$ Pa s |
| phenyl-aceline | $\langle\tau_\varepsilon\rangle$ | 142 | 19.0 | $9.5 \times 10^{-16}$ s |
| | $\rho_{\text{dc}}$ | 97 | 65.9 | $7.6 \times 10^{-11}$ $\Omega^{-1}$ cm$^{-1}$ |
| | $\eta$ | / | / | / |

**TABLE 1.** The parameters obtained from VFT fits [Eq. (3)] of $\langle\tau_\varepsilon\rangle(T)$, $\rho(T)$, and $\eta(T)$ for oxaline, maline, and phenylaceline. For phenylaceline, the parameters of the fit of $\langle\tau_\varepsilon\rangle$ accounting for $T_g^{\text{DSC}}$ [dashed line in Fig. 4(b)] are shown.

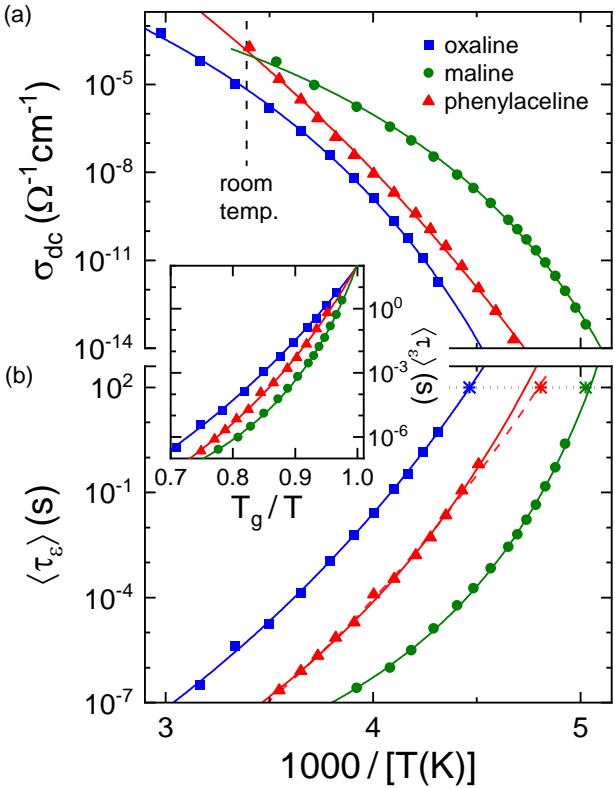

**FIG. 4.** Arrhenius representation of the dc conductivity $\sigma_{dc}$ (a) and average dipolar relaxation time $\langle\tau_\varepsilon\rangle$ (b), as obtained from fits of the dielectric spectra for maline, oxaline, and phenylaceline. The stars show the respective glass-transition temperatures $T_g^{DSC}$ as determined from DSC measurements, assuming $\langle\tau\rangle(T_g) \approx 100$ s. The solid lines represent VFT fits [Eq. (3)] with the parameters listed in Table 1. The dashed line in (b) is an alternative VFT fit of $\langle\tau_\varepsilon\rangle$ of phenylaceline, also considering the DSC data point. The inset shows the average dipolar relaxation times of the three carboxylic acid-based DESs in an Angell plot.

Figure 4(a) reveals details regarding the ionic transport of the investigated DESs. The room-temperature dc conductivity is the most important quantity in terms of application. For maline and phenylaceline it is close to the technically relevant value of $\sigma_{dc} > 10^{-4}$ $\Omega^{-1}$ cm$^{-1}$,[38,46,47] while for oxaline it is smaller. In general, the detected conductivities are somewhat lower than those reported in Ref. 1, which we attribute to differences in the syntheses. The heating method, as employed in that work, can result in the formation of esters and water.[24,59] In particular, the latter can significantly modify the physiochemical properties of DESs,[21,101,102,103] e.g., the presence of water can enhance the dc conductivity across the entire temperature range.[53,104,105] Hence, we avoided heating as much as possible during our synthesis process, leading to lower intrinsic dc conductivities in the final product.

Phenylaceline displays a higher $\sigma_{dc}$ compared to oxaline and a similar room-temperature value in comparison to maline. Additionally, phenylaceline exhibits a higher $T_{VF}^\sigma$ than maline which is the main reason why it reveals a strongly reduced dc conductivity at low temperatures. In addition, the fragility of phenylaceline is significantly lower than that of maline (i.e., its $D_\sigma$ is larger, cf. Table 1), which results in a less curved $\sigma_{dc}(1/T)$-trace for phenylaceline in Fig. 4(a) and rationalizes why the two liquids reveal similar conductivities at room temperature although their Vogel-Fulcher temperatures are different. This finding exemplifies how the interplay of parameters related to the low-temperature glass formation significantly impact the physical properties of DESs, even at room temperature.

At first glance, in Fig. 4, the dc conductivity of each of the three DESs seems to roughly follow a temperature profile that is inversely proportional to the average relaxation time. Figure 5 further analyses the relation of the two quantities. It displays Arrhenius plots of the dielectric and mechanical dynamic quantities acquired from the fits for each of the three DESs investigated, namely the average dielectric and rheological relaxation times ($\langle\tau_\varepsilon\rangle$ and $\langle\tau_J\rangle$, respectively), the dc resistivity, $\rho_{dc} = 1/\sigma_{dc}$, and the shear viscosity $\eta$. In Fig. 5, each ordinate covers the same number of decades which facilitates the direct comparison of the different quantities. As mentioned, mechanical data could not be obtained for phenylaceline. Additionally, the glass-transition temperatures $T_g^{DSC}$ derived from DSC measurements are included, presuming the common definition $\langle\tau_\varepsilon\rangle(T_g) \approx 100$ s. The ordinates for the three quantities were adjusted to ensure that the expected values at $T_g$ [$\tau(T_g) = 100$ s, $\eta(T_g) = 10^{12}$ Pa s, and $\rho_{dc}(T_g) = 10^{15} \Omega$ cm] appear at the same level on the y axis. This enables a convenient comparison of the different dynamic quantities and an assessment of their possible decoupling. In the case of $\rho_{dc}$, an approach adopted from Mizuno et al.[106] is applied: Assuming that the translational motion rates of the conducting particles match the frequency of the relaxation, the Stokes-Einstein and Nernst-Einstein relations lead to a dc resistivity of approximately $10^{15}$ $\Omega$ cm at the glass-transition temperature.

Interestingly, at $T_g^{DSC}$ in oxaline and maline, $\langle\tau_J\rangle$, $\rho_{dc}$, and $\eta$ do not align with their expected values, in contrast to $\langle\tau_\varepsilon\rangle$ which exhibits a close match. This result suggests that in these materials the glass transition detected using DSC is mainly determined by the orientational freezing of the dipolar constituents. This finding corresponds to observations in supercooled plastic crystals[107,108,109,110] where only orientational motions are possible. Similarly, for some conventional structural glass formers such as amorphous ethanol,[111,112] it was found that the freezing of the rotational degrees of freedom dominates the glass transition and that flow processes are less important. In contrast to $\langle\tau_\varepsilon\rangle$, the other displayed quantities are also sensitive to the dynamics of the non-polar DES constituents. In oxaline and maline their degrees of freedom seem to freeze in at lower temperatures upon cooling, with limited involvement in the "real" glass-transition detected by DSC. For phenylaceline, the situation is different: An extrapolated VFT fit of the dc resistivity reasonably matches the anticipated value of $10^{15}$ Ω cm at $T_g^{DSC}$, within experimental uncertainty. In phenylaceline, the ionic translational diffusion seems to be more closely linked to the glassy freezing than it is for oxaline and maline.

Figures 5(a) and (b) provide valuable insights not only into the glass transition of oxaline and maline but also into their translational dynamics. The average relaxation time for $\langle\tau_J\rangle$, as derived from the shear compliance (plusses), exhibits a temperature dependence similar to that of $\langle\tau_\varepsilon\rangle$ (crosses) while displaying significantly smaller absolute values. Within experimental uncertainty, for oxaline, the two time-scales constantly differ by approximately 1.2 decades. This also becomes evident from the temperature independence of the decoupling index $\log_{10}(\langle\tau_\varepsilon\rangle/\langle\tau_J\rangle)$ [inset of Fig. 5(b)], which is similar to measures considered by other authors.[113,114,115] For maline, the decoupling index significantly increases with decreasing temperature. This finding further highlights the faster freezing of the rotational motions of maline's dipolar species in comparison to the more gradual slowing down of the entire liquid's flow dynamics reflected by $\langle\tau_J\rangle$. The difference in absolute $\tau$ values may be attributed to the distinct components that are investigated by each measurement technique: While DS only tracks the large, dipolar components of the DESs, rheological measurements also sense the dynamics of the smaller and faster[116] chloride-anions, resulting in generally shorter average structural relaxation times.

The viscosity is governed by the same dynamics which controls the shear-stress relaxation. This statement is fully confirmed by the equivalent temperature dependences of $\langle\tau_J\rangle$ and $\eta$ in the rheologically tested temperature range of oxaline and maline [Figs. 5(a) and (b)]. A more thorough evaluation of the structural dynamics over a wider temperature range can be achieved by considering high-temperature viscosity data from literature.[24] In fact, for oxaline at high temperatures, the difference between the $\langle\tau_\varepsilon\rangle(1/T)$ and $\eta(1/T)$ traces in Fig. 5(a) significantly changes with temperature. This differs from the constant decoupling between $\langle\tau_\varepsilon\rangle$ and $\langle\tau_J\rangle$ reported in the inset of Fig. 5 for the low-temperature range. As revealed by Table 1, the strength parameters $D$ and the divergence temperatures $T_{VF}$ relating to $\langle\tau_\varepsilon\rangle$, indeed, differ significantly from those relating to $\eta$, indicating an only limited correlation between rotational dipolar and predominantly translational structural dynamics. A similar dynamical decoupling is noticeable for maline near $T_g$ [Fig. 5(b) and inset]. As discussed, e.g., in Ref. 55, the Debye-Stokes-Einstein (DSE) relation predicting $D_r \propto T/\eta$ (with $D_r$ the rotational diffusion coefficient)[117] essentially implies the proportionality $\tau_\varepsilon \sim \eta$. However, just as for the present DESs, a breakdown of this relation is often observed in supercooled liquids near $T_g$ which

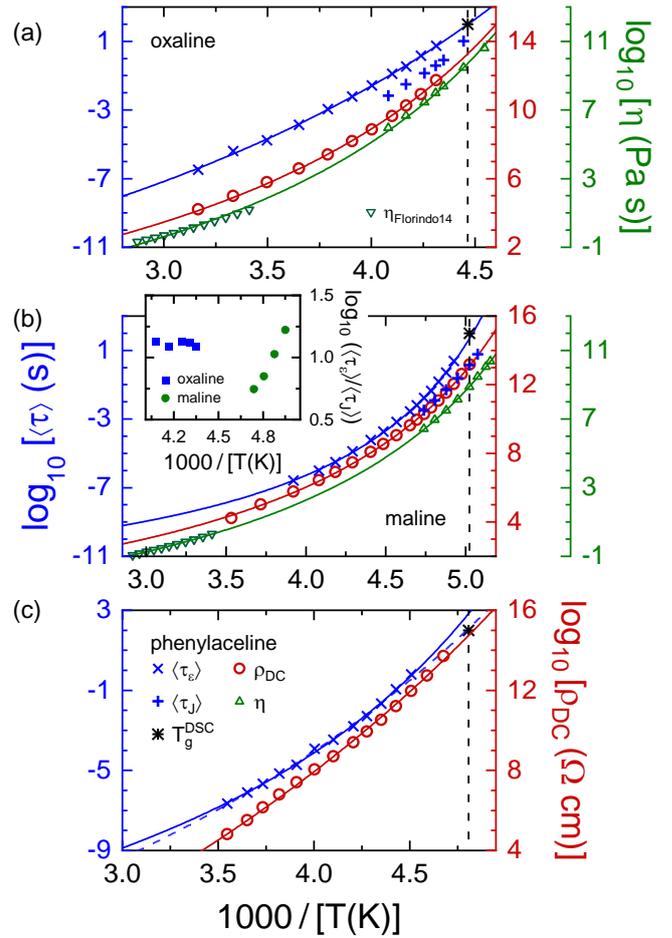

**FIG. 5.** Arrhenius plots of different dynamic quantities for oxaline (a), maline (b), and phenylaceline (c). Plotted are the average dipolar relaxation time $\langle\tau_\varepsilon\rangle$ (crosses, left ordinate), the dc resistivity $\rho_{dc} = 1/\sigma_{dc}$ (circles, right ordinate), the average relaxation time $\langle\tau_J\rangle$ (pluses, left ordinate), and the viscosity $\eta$ (upward triangles, right ordinate, not available for phenylaceline) as obtained from fits of the dielectric and rheological spectra. The different ordinates cover the same number of decades. Additionally, the glass-transition temperature $T_g^{DSC}$ (stars), determined from DSC measurements, and the viscosity $\eta_{Florindo14}$ (downward triangles, rightmost ordinate), reported in Ref. 24, are displayed. The ordinates for the three quantities were adjusted to ensure that the expected values at $T_g$ ($\tau = 100$ s, $\eta = 10^{12}$ Pa s, $\rho_{dc} = 10^{15}$ Ω cm) appear at the same level. The solid lines represent VFT fits [Eq. (3)] of the respective quantities with the corresponding parameters listed in Table 1. The VFT fits of $\eta$ take the literature data into account. The dashed line in (c) presents an alternative VFT fit of $\langle\tau_\varepsilon\rangle$ accounting for $T_g^{DSC}$. The inset in (b) features an Arrhenius representation of the decoupling index $\log(\langle\tau_\varepsilon\rangle/\langle\tau_J\rangle)$ for oxaline and maline.

is believed to be an inherent characteristic of the liquid state.[118,119,120,121,122,123,124,125]

It should be noted that for oxaline the two $\eta$ datasets cannot be precisely aligned due to their slightly different curvatures [Fig. 5(a)]. This discrepancy is especially prominent in the high-temperature region when compared to the VFT fit of both data sets. This minor, yet evident inconsistency probably hints at slightly different water concentrations in the samples. As already mentioned, the presence of water in DESs has a notable impact on their physicochemical properties, particularly their viscosity.[24,53,104,126,127,128] Here, it seems that water does not play a decisive role with regards to the observed quantities and that it may be disregarded for the remainder of this study.

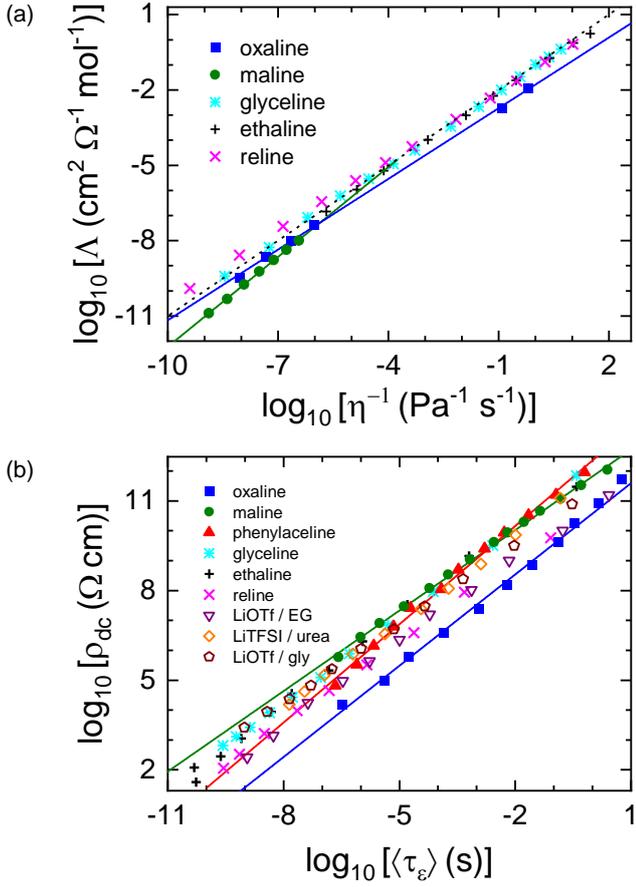

**FIG. 6.** (a) Walden plot displaying the molar conductivity $\Lambda = \sigma_{dc}/c_{ion}$ against the fluidity $\eta^{-1}$ of oxaline and maline in a double-logarithmic presentation, compared to previously investigated ChCl-based DESs.[55] The viscosities at the temperatures for which conductivity data are available were determined by a B-spline interpolation of $\eta(1000/T)$. To calculate the ion concentration $c_{ion}$ of the DESs, the densities obtained by Wang *et al.*[131] at 25°C were used. The solid lines with slope $\xi = 0.94$ and 1.19 show fits for oxaline and maline, respectively. The dashed line represents an ideal Walden behavior, $\Lambda \sim \eta^{-\xi}$ with $\xi = 1$. (b) Dc resistivity displayed against $\langle \tau_\varepsilon \rangle$ for all three investigated DESs, together with data for previously investigated ChCl-based[55] and Li-salt-based DESs.[56] The solid lines highlight that the dc resistivity of the investigated electrolytes can be described by a power law $\rho_{dc} \sim \langle \tau_\varepsilon \rangle^\xi$ with $0.89 < \xi < 1.09$. The line for oxaline corresponds to $\xi = 1$.

Finally, we evaluate the dc resistivity with the goal to characterize the translational ion motion of the DESs. In oxaline as well as in maline, the temperature dependence of $\rho_{dc}$ matches the temperature profile of $\eta$ across the entire observed temperature range [Figs. 5(a) and (b)], and consequently the VFT fit parameters are similar (Table 1). Figure 6(a) explores the relation between the ionic and the structural transport dynamics in terms of a Walden plot of the molar conductivity $\Lambda = \sigma_{dc}/c_{ion}$, where $c_{ion}$ represents the molar ion concentration, against the inverse shear viscosity.[129,130] For comparison, also data on glyceline, ethaline, reline, and three Li-salt-based DESs are included in this figure, based on data reported in Refs. 55 and 56. Considering the small thermal expansion coefficient of the investigated DESs,[25] the ion concentration $c_{ion}$ is determined in good approximation using densities obtained at 25°C from Ref. 131.

The Walden rule, $\Lambda\eta = $ const., can be derived by combining the Stokes-Einstein (SE) relation ($D_t \sim T/\eta$, with $D_t$ denoting the translational diffusion coefficient) and the Nernst-Einstein relation ($\sigma_{dc} \sim D_i/T$, with $D_i$ designating the ionic diffusion coefficient), by assuming $D_i = D_t$.[132] In principle, the Walden rule should apply to DESs, since it was originally formulated for solutions of ions in a polar liquid.[133] However, Fig. 6(a) shows that for oxaline and for maline, the molar conductivity can be better described by a power law $\Lambda \sim \eta^{-\xi}$.[134] With a slope of $\xi = 0.94$, oxaline deviates only slightly from the Walden rule and can still be considered a reasonably good Walden electrolyte.[135] Thus, the fragility and glass-transition temperature, which determine the viscosity, have a direct impact on its conductivity, even at room-temperature, which is a frequently neglected aspect of glass-forming ionic conductors. Oxaline shows somewhat larger deviations from ideal Walden behavior as compared to the previously investigated ChCl-based DESs ethaline and glyceline, which reveal a slope of one in Fig. 6(a) (dashed line).[55] Its somewhat lower molar conductivity for a given viscosity may be attributed to the strong hydrogen bonds between $Cl^-$ and the oxalic-acid molecules, resulting in a decreased mobility of the anion,[97] as well as an enhanced charge spreading between the ions and the HBD.[96] Although a similar situation is expected for maline, a clear super-linear dependence, with $\xi = 1.19$, becomes evident for this DES in Fig. 6(a). Unlike reline[55] and several other DESs,[131] which also exhibit a non-trivial relation of ion mobility and viscosity, maline displays a slope larger than unity in the high-viscosity area of the Walden plot. In general, the decoupling of translational and rotational dynamics, as well as the clear violation of the Stokes-Einstein relation near $T_g$, are well-known phenomena in glass-forming materials.[115,119,122,123,124,119,136,137,138,139]

Lastly, Fig. 6(b) displays the dc resistivity versus the average dipolar relaxation time for the investigated DESs in a double-logarithmic presentation, and compares these data with those for the DESs examined in Refs. 54-56. It is noteworthy that oxaline displays the lowest dc resistivity and, hence, the highest dc conductivity for a given dipolar relaxation time. As revealed by Fig. 4(a), this feature is not a result of an exceptionally high ionic conductivity but rather can be attributed to the exceptionally slow reorientation of the dipoles. A direct proportionality between the ionic charge transport and the dipolar reorientation dynamics exists for oxaline, as confirmed by the fit with slope one [(solid blue line in Fig. 6(b)] of the respective logarithmic quantities, implying $\rho_{dc} \sim \tau_\varepsilon$. It should be noted that such a proportionality sometimes also is denoted as DSE relation.[140,141] Based on the only slight deviations of oxaline from the Walden rule [Fig. 6(a)], the translational dynamics of the ions in this DES is primarily driven by the viscosity and not directly affected by the reorientational dynamics of the dipoles as considered in revolving-door like charge transport mechanisms.[46,54,73,142] The proportionality $\rho_{dc} \propto \tau_\varepsilon$ revealed in Fig. 6(b) then is due to an indirect coupling of both quantities via the viscosity. This finding aligns with previous observations made for ethaline and glyceline.[55,143]

However, the situation in maline and phenylaceline is more complex. Here, the relationship between the dc resistivity and the dipolar relaxation time is better described by a power law $\rho_{dc} \sim \langle \tau_\varepsilon \rangle^\xi$ with $\xi = 0.89$ and $\xi = 1.09$, respectively. For $\xi < 1$, such a behavior, sometimes termed fractional DSE relation, was previously reported for various glass formers.[140,141] An exponent $\xi \neq 1$ indicates a decoupling of the ionic translational and dipolar rotational dynamics. If one assumes a full coupling at high temperatures, an exponential factor $\xi < 1$, as found in maline, reline, and the previously studied Li-salt-based DESs,[54,56] indicates the enhancement of the ionic conductivity *for a given relaxation time* when compared to electrolytes for which both quantities are fully coupled. Such enhancement should not be confused with an increase in $\sigma_{dc}$ *at a given temperature*. An exponent $\xi > 1$, such as found for phenylaceline describes the opposite effect. We are only aware of one instance in the literature on DESs where a similar behavior has been reported.[68] Overall, the decoupling of the two dynamics undoubtedly affects the ionic conductivity in both systems, albeit in distinct manners.

## IV. SUMMARY AND CONCLUSIONS

In summary, we have performed dielectric spectroscopy and shear rheological measurements in a broad temperature and dynamic range on three carboxylic-acid-based NADESs: oxaline, maline, and phenylaceline. The dielectric permittivity spectra allow for the characterization of the translational ionic and reorientational dipolar dynamics by means of dc conductivity and dipolar relaxation time, respectively. Complementarily, the mechanical compliance spectra provide insights into the flow behavior by accessing viscosity and structural relaxation times.

All of the three carboxylic-acid-based DESs reveal characteristics that indicate glassy freezing. Especially, their temperature-dependent dynamics display significant deviations from a simple, thermally activated Arrhenius behavior. Instead, the dipolar relaxation time, viscosity, and dc conductivity are more accurately described using the VFT equation. In oxaline and maline, the glass transition as detected by DSC measurements is governed by the orientational degrees of freedom of the dipolar constituents rather than by flow processes. In all three materials, the glassy freezing has a significant impact on their ionic conductivity, even at room temperature. A closer examination reveals complex relationships and varying degrees of decoupling between the different, observed dynamics in all three systems. Among them, oxaline features the smallest deviations, with dipolar reorientational and ionic translational motions showing an apparent interdependence. For this DES, only small deviations from a direct proportionality between molar conductance and shear viscosity were observed. Maline exhibits notable decoupling effects that are evidenced by a super-linear Walden rule and, at low temperatures, an enhanced ionic conductivity for a given dipolar relaxation time when compared to fully coupled electrolytes. Both these decoupling effects are most pronounced near the glass-transition temperature. In phenylaceline, the dc resistivity increases super-linearly with the dipolar relaxation time, revealing a decoupling effect opposite to that found for maline.

As demonstrated in this study, the choice of the HBD bears a major impact on the molecular and ionic dynamics of the DESs. To determine HBD/HBA pairings that optimally exhibit the desired physiochemical properties for a specific application, a detailed understanding of the complex interrelationships of the constituents is essential. Furthermore, it is crucial not to ignore the aspects related to the glass formation of DESs, in particular their effect on the ionic conductivity, even at room temperature, which is important for potential applications as electrolytes.

## ACKNOWLEDGMENTS

This work was supported by the Deutsche Forschungsgemeinschaft (project no. 444797029).

## AUTHOR DECLARATIONS

### Conflict of Interest

The authors have no conflicts to disclose.

### DATA AVAILABILITY

The data that support the findings of this study are available from the corresponding author upon reasonable request.

———————————————

## REFERENCES

[1] A. P. Abbott, D. Boothby, G. Capper, D. L. Davies, and R. K. Rasheed, J. Am. Chem. Soc. **126**, 9142 (2004).
[2] B. B. Hansen, S. Spittle, B. Chen, D. Poe, Y. Zhang, J. M. Klein, A. Horton, L. Adhikari, T. Zelovich, B. W. Doherty, B. Gurkan, E. J. Maginn, A. Ragauskas, M. Dadmun, T. A. Zawodzinski, G. A. Baker, M. E. Tuckerman, R. F. Savinell, and J. R. Sangoro, Chem. Rev. **121**, 1232 (2020).
[3] T. E. Achkar, H. Greige-Gerges, and S. Fourmentin, Environ. Chem. Lett. **19**, 3397 (2021).
[4] H. Qin, X. Hu, J. Wang, H. Cheng, L. Chen, and Z. Qi, Green Energy Environ. **5**, 8 (2020).
[5] F. Pena-Pereira and J. Namieśnik, ChemSusChem. **7**, 1784 (2014).
[6] C. Florindo, L. C. Branco, and I. M. Marrucho, ChemSusChem **12**, 1549 (2019).
[7] D. Yu, Z. Xue and T. Mu, Chem. Soc. Rev. **50**, 8596 (2021).
[8] L. I. N. Tomé, V. Baião, W. da Silva, and C. M. A. Brett, Appl. Mater. Today **10**, 30 (2018).
[9] E. L. Smith, A. P. Abbott, and K. S. Ryder, Chem. Rev. **114**, 11060 (2014).
[10] Q. Zhang, K. D. O. Vigier, S. Royer, and F. Jérôme, Chem. Soc. Rev. **41**, 7108 (2012).
[11] M. A. R. Martins, S. P. Pinho, and J. A. P. Coutinho, J. Solution. Chem. **48**, 962 (2018).

[12] A. P. Abbott, G. Capper, D. L. Davies, H. L. Munro, R. K. Rasheed, and V. Tambyrajah, Chem. Commun. **2010**, 2001.
[13] S. L. Perkins, P. Painter, and C. M. Colina, J. Chem. Eng. Data **59**, 3652 (2014).
[14] O. S. Hammond, D. T. Bowron, and K. J. Edler, Green Chem. **18**, 2736 (2016).
[15] M. Francisco, A. van den Bruinhorst, and M. C. Kroon, Green Chem. **14**, 2153 (2012).
[16] A. Paiva, R. Craveiro, I. Aroso, M. Martins, R. L. Reis, and A. R. C. Duarte, ACS Sustain. Chem. Eng. **2**, 1063 (2014).
[17] K. A. Omar and R. Sadeghi, J. Chem. Eng. Data **66**, 2088 (2021).
[18] S. Tang, G. A. Baker, and H. Zhao, Chem. Soc. Rev. **41**, 4030 (2012).
[19] Y. Dai, J. van Spronsen, G.-J. Witkamp, R. Verpoorte, and Y. H. Choi, Anal. Chim. Acta. **766**, 61 (2013).
[20] R. E. Owyeung, S. R. Sonkusale, and M. J. Panzer, J. Phys. Chem. B **124**, 5986 (2020).
[21] Y. Dai, G.-J. Witkamp, R. Verpoorte, and Y. H. Choi, Food Chem. **187**, 14 (2015).
[22] B. Singh, H. Lobo, and G. Shankarling, Catal. Lett. **141**, 178 (2010).
[23] J. Płotka-Wasylka, M. de la Guardia, V. Andruch, and M. Vilková, Microchem. J. **159**, 105539 (2020).
[24] C. Florindo, A. M. Fernandes, and I. M. Marrucho, ACS Sustain. Chem. Eng. **2**, 2416 (2014).
[25] G. García, S. Aparicio, R. Ullah, and M. Atilhan, Energy Fuels **29**, 2616 (2015).
[26] Y. Dai, G.-J. Witkamp, R. Verpoorte, and Y. H. Choi, Anal. Chem. **85**, 6272 (2013).
[27] K. Radošević, M. C. Bubalo, V. G. Srček, D. Grgas, T. L. Dragičević, and I. R. Redovniković, Ecotoxicol. Environ. Saf. **112**, 46 (2015).
[28] M. Espino, M. de los Ángeles Fernández, F. J. V. Gomez, and M. F. Silva, Trends Analyt. Chem. **76**, 126 (2016).
[29] M. Francisco, A. van den Bruinhorst, and M. C. Kroon, Angew. Chem. Int. Ed. **52**, 3074 (2013).
[30] P. T. Anastas and T. C. Williamson, ACS Symp. Ser. **626**, 1 (1996).
[31] P. T. Anastas and M. M. Kirchhoff, Acc. Chem. Res. **35**, 686 (2002).
[32] K. Grodowska and A. Parczewski, Acta Pol. Pharm. **67**, 3 (2010).
[33] D. Prat, O. Pardigon, H.-W. Flemming, S. Letestu, V. Ducandas, P. Isnard, E. Guntrum, T. Senac, S. Ruisseau, P. Cruciani, and P. Hosek, Org. Process. Res. Dev. **17**, 1517 (2013).
[34] X. Liu, S. Ahlgren, H. A. A. J. Korthout, L. F. Salomé-Abarca, L. M. Bayona, R. Verpoorte, and Y. H. Choi, J. Chromatogr. A **1532**, 198 (2018).
[35] Q. Zhang, Q. Wang, S. Zhang, X. Lu, and X. Zhang, ChemPhysChem **17**, 335 (2015).
[36] H. Cruz, N. Jordão and L. C. Branco, Green Chem. **19**, 1653 (2017).
[37] R. Wang, S. Zhang, Y. Su, J. Liu, Y. Ying, F. Wang, and X. Cao, Electrochim. Acta. **258**, 328 (2017).
[38] H. Cruz, N. Jordao, A. L. Pinto, M. Dionisio, L. A. Neves, and L. C. Branco, ACS Sustain. Chem. Eng. **2**, 222 (2020).
[39] Y. Alesanco, A. Viñuales, J. Rodriguez, and R. Tena-Zaera, Materials **11**, 414 (2018).
[40] D. Larcher and J.-M. Tarascon, Nat. Chem. **7**, 19 (2014).
[41] M. Armand and J.-M. Tarascon, Nature **451**, 652 (2008).
[42] Z. Zhang and L. F. Nazar, Nat. Rev. Mater. **7**, 389 (2022).
[43] H.-R. Jhong, D. S.-H. Wong, C.-C. Wan, Y.-Y. Wang, and T.-C. Wei, Electrochem. Commun. **11**, 209 (2009).
[44] D. D. Marino, M. Shalaby, S. Kriescher, and M. Wessling, Electrochem. Commun. **90**, 101 (2018).
[45] A. H. Whitehead, M. Pölzler, and B. Gollas, J. Electrochem. Soc. **157**, D328 (2010).
[46] D. R. MacFarlane and M. Forsyth, Adv. Mater. **13**, 957 (2001).
[47] W. Lu, A. G. Fadeev, B. Qi, E. Smela, B. R. Mattes, J. Ding, G. M. Spinks, J. Mazurkiewicz, D. Zhou, G. G. Wallace, D. R. MacFarlane, S. A. Forsyth, and M. Forsyth, Science **297**, 983 (2002).
[48] O. E. Geiculescu, D. D. DesMarteau, S. E. Creager, O. Haik, D. Hirshberg, Y. Shilina, E. Zinigrad, M. D. Levi, D. Aurbach, and I. C. Halalay, J. Power Sources **307**, 519 (2016).
[49] S. Cai, X. Chu, C. Liu, H. Lai, H. Chen, Y. Jiang, F. Guo, Z. Xu, C. Wang, and C. Gao, Adv. Mater. **33**, 2007470 (2021).
[50] N. Tekin, M. Cebe, and Ç. Tarımcı, Chem. Phys. **300**, 239 (2004).
[51] M. Samsonowicz, Spectrochim. Acta A **118**, 1086 (2014).
[52] E. Vilaseca, E. Perez, and F. Mata, Chem. Phys. Lett. **132**, 305 (1986).
[53] A. Jani, B. Malfait, and D. Morineau, J. Chem. Phys. **154**, 164508 (2021).
[54] D. Reuter, C. Binder, P. Lunkenheimer, and A. Loidl, Phys. Chem. Chem. Phys. **21**, 6801 (2019).
[55] D. Reuter, P. Münzner, C. Gainaru, P. Lunkenheimer, A. Loidl, and R. Böhmer, J. Chem. Phys. **154**, 154501 (2021).
[56] A. Schulz, P. Lunkenheimer, and A. Loidl, J. Chem. Phys. **155**, 044503 (2021).
[57] S. Lansab, T. Schwan, K. Moch, and R. Böhmer, J. Chem. Phys. (submitted Nov. 2023).
[58] S. Yasmin, W.-B. Sheng, C.-Y. Peng, A.-U. Rahman, D.-F. Liao, M. I. Choudhary, and W. Wanga, Synth. Commun. **48**, 68 (2017).
[59] N. R. Rodriguez, A. van den Bruinhorst, L. J. B. M. Kollau, M. C. Kroon, and K. Binnemans, ACS Sustain. Chem. Eng. **7**, 11521 (2019).
[60] V. Agieienko and R. Buchner, Phys. Chem. Chem. Phys. **22**, 20466 (2020).
[61] V. Agieienko and R. Buchner, J. Chem. Eng. Data **64**, 4763 (2019).
[62] R. Böhmer, M. Maglione, P. Lunkenheimer, and A. Loidl, J. Appl. Phys. **65**, 901 (1989).
[63] P. Lunkenheimer, V. Bobnar, A. V. Pronin, A. I. Ritus, A. A. Volkov, and A. Loidl, Phys. Rev. B **66**, 052105 (2002).
[64] P. Lunkenheimer, S. Krohns, S. Riegg, S. G. Ebbinghaus, A. Reller, and A. Loidl, Eur. Phys. J. Spec. Top. **180**, 61 (2010).
[65] S. Emmert, M. Wolf, R. Gulich, S. Krohns, S. Kastner, P. Lunkenheimer, and A. Loidl, Eur. Phys. J. B **83**, 157 (2011).
[66] T. Wang, J. Hu, H. Yang, L. Jin, X. Wei, C. Li, F. Yan, and Y. Lin, J. Appl. Phys. **121**, 084103 (2017).
[67] A. Serghei, M. Tress, J. R. Sangoro, and F. Kremer, Phys. Rev. B **80**, 184301 (2009).
[68] C. D'Hondt and D. Morineau, J. Mol. Liq. **365**, 120145 (2022).
[69] S. Spittle, D. Poe, B. Doherty, C. Kolodziej, L. Heroux, M. A. Haque, H. Squire, T. Cosby, Y. Zhang, C. Fraenza, S. Bhattacharyya, M. Tyagi, J. Peng, R. A. Elgammal, T. Zawodzinski, M. Tuckerman, S. Greenbaum, B. Gurkan, C. Burda, M. Dadmun, E. J. Maginn, and J. Sangoro, Nat. Commun. **13**, 219 (2022).
[70] A. Schulz, P. Lunkenheimer, and A. Loidl, arXiv:2311.09360 (2023).
[71] C. Krause, J. R. Sangoro, C. Iacob, and F. Kremer, J. Phys. Chem. B **114**, 382 (2009).
[72] P. Sippel, P. Lunkenheimer, S. Krohns, E. Thoms, and A. Loidl, Sci. Rep. **5**, 13922 (2015).
[73] P. Sippel, S. Krohns, D. Reuter, P. Lunkenheimer, and A. Loidl, Phys. Rev. E **98**, 052605 (2018).
[74] G. P. Johari and M. Goldstein, J. Chem. Phys. **53**, 2372 (1970).
[75] M. D. Ediger, C. A. Angell, and S. R. Nagel, J. Phys. Chem. **100**, 13200 (1996).
[76] J. C. Dyre, J. Appl. Phys. **64**, 2456 (1988).
[77] T. B. Schrøder and J. C. Dyre, Phys. Rev. Lett. **101**, 025901 (2008).
[78] K. Funke, Prog. Solid State Chem. **22**, 111 (1993).
[79] K. S. Cole and R. H. Cole, J. Chem. Phys. **9**, 341 (1941).
[80] S. Havriliak, S. Negami, J. Polymer Sci. C **14**, 99 (1966).
[81] D. W. Davidson and R. H. Cole, J. Chem. Phys. **19**, 1484 (1951).
[82] S. P. Bierwirth, R. Böhmer, and C. Gainaru, Phys. Rev. Lett. **119** (2017).

[83] P. Münzner, L. Hoffmann, R. Böhmer, and C. Gainaru, J. Chem. Phys. **150**, 234505 (2019).
[84] L. Röwekamp, K. Moch, C. Gainaru, and R. Böhmer, Mol. Pharmaceutics **19**, 1586 (2022).
[85] The supplementary information of S. P. Bierwirth, C. Gainaru, and R. Böhmer, J. Chem. Phys. **148**, 221102 (2018) describes the procedure that we apply to obtain rheological master curves in detail.
[86] J. C. Dyre, Rev. Mod. Phys. **78**, 953 (2006).
[87] H. Vogel, Phys. Z. **22**, 645 (1921).
[88] G. S. Fulcher, J. Am. Ceram. Soc. **8**, 339 (1925).
[89] G. Tammann and W. Hesse, Z. Anorg. Allg. Chem. **156**, 245 (1926).
[90] C. A. Angell, in *Relaxations in Complex Systems*, edited by K. L. Ngai and G. B. Wright (NRL, Washington, DC, 1985), p. 3.
[91] R. Böhmer and C. A. Angell, Phys. Rev. B **45**, 10091 (1992).
[92] R. Böhmer, K. L. Ngai, C. A. Angell, and D. J. Plazek, J. Chem. Phys. **99**, 4201 (1993).
[93] C. A. Angell, J. Phys. Chem. Solids **49**, 863 (1988).
[94] D. V. Wagle, C. A. Deakyne, and G. A. Baker, J. Phys. Chem. B **120**, 6739 (2016).
[95] S. Sahu, S. Banu, A. K. Sahu, B. V. N. P. Kumar, and A. K. Mishra, J. Mol. Liq. **350**, 118478 (2022).
[96] S. Zahn, B. Kirchner, and D. Mollenhauer, ChemPhysChem **17**, 3354 (2016).
[97] Z. Naseem, R. A. Shehzad, A. Ihsan, J. Iqbal, M. Zahid, A. Pervaiz, and G. Sarwari, Chem. Phys. Lett. **769**, 138427 (2021).
[98] Y. M. Jung, Bull Korean Chem Soc. **24**, 1410 (2003).
[99] M. N. Levandowski, T. C. Rozada, U. Z. Melo, E. A. Basso, and B. C. Fiorin, Spectrochim. Acta A **174**, 138 (2017).
[100] E. M. S. Maçôas, R. Fausto, J. Lundell, M. Pettersson, L. Khriachtchev, and M. Räsänen, J. Phys. Chem. A **104**, 11725 (2000).
[101] C. Ma, A. Laaksonen, C. Liu, X. Lu, and X. Ji, Chem. Soc. Rev. **47**, 8685 (2018).
[102] P. Kalhor, Y.-Z. Zheng, H. Ashraf, B. Cao, and Z.-W. Yu, ChemPhysChem **21**, 995 (2020).
[103] S. Kaur, A. Gupta, and H. K. Kashyap, J. Phys. Chem. B **124**, 2230 (2020).
[104] C. Du, B. Zhao, X.-B. Chen, N. Birbilis, and H. Yang, Sci. Rep. **6**, 29225 (2016).
[105] L. Gontrani, M. Bonomo, N. V. Plechkova, D. Dini, and R. Caminiti, Phys. Chem. Chem. Phys. **20**, 30120 (2018).
[106] F. Mizuno, J.-P. Belieres, N. Kuwata, A. Pradel, M. Ribes, and C. A. Angell, J. Non-Cryst. Solids **352**, 5147 (2006).
[107] A. Loidl and R. Böhmer, in: *Disorder effects on relaxational processes*, edited by R. Richert and A. Blumen (Springer, Berlin, 1994), p. 659.
[108] R. Brand, P. Lunkenheimer, and A. Loidl, J. Chem. Phys. **116**, 10386 (2002).
[109] L. C. Pardo, P. Lunkenheimer, and A. Loidl, J. Chem. Phys. **124**, 124911 (2006).
[110] L. P. Singh and S. S. N. Murthy, Phys. Chem. Chem. Phys. **11**, 5110 (2009).
[111] M. A. Ramos, S. Vieira, F. J. Bermejo, J. Dawidowski, H. E. Fischer, H. Schober, M. A. González, C. K. Loong, and D. L. Price, Phys. Rev. Lett. 78, **82** (1997).
[112] M. A. Ramos, I. M. Shmyt'ko, E. A. Arnautova, R. J. Jiménez-Riobóo, V. Rodríguez-Mora, S. Vieira. and M. J. Capitán, J. Non-Cryst. Solids **352**, 4769 (2006).
[113] C. A. Angell, Chem. Rev. **90**, 523 (1990).
[114] R. Zorn, F. I. Mopsik, G. B. McKenna, L. Willner, and D. Richter, J. Chem. Phys. **107**, 3645 (1997).
[115] I. Chang and H. Sillescu, J. Phys. Chem. B **101**, 8794 (1997).
[116] Y. Hinz, J. Beerwerth, and R. Böhmer, Phys. Chem. Chem. Phys. **25**, 28130 (2023)
[117] P. Debye, Polar Molecules (Dover Publications, 1929).
[118] E. Rössler, Phys. Rev. Lett. **65**, 1595 (1990).
[119] F. Fujara, B. Geil, H. Sillescu, and G. Fleischer, Z. Phys. B **88**, 195 (1992).
[120] G. Tarjus and D. Kivelson, J. Chem. Phys. **103**, 3071 (1995).
[121] L. Andreozzi, A. D. Schino, M. Giordano, and D. Leporini, J. Phys.: Condens. Matter **8**, 9605 (1996).
[122] K. L. Ngai, Philos. Mag. B **79**, 1783 (1999).
[123] S. R. Becker, P. H. Poole, and F. W. Starr, Phys. Rev. Lett. **97**, 055901 (2006).
[124] M. G. Mazza, N. Giovambattista, H. E. Stanley, and F. W. Starr, Phys. Rev. E **76**, 031203 (2007).
[125] J. Singh and P. P. Jose, J. Phys.: Condens. Matter **33**, 055401 (2020).
[126] T. E. Achkar, S. Fourmentin, and H. Greige-Gerges, J. Mol. Liq. **288**, 111028 (2019).
[127] G. C. Dugoni, M. E. D. Pietro, M. Ferro, F. Castiglione, S. Ruellan, T. Moufawad, L. Moura, M. F. C. Gomes, S. Fourmentin, and A. Mele, ACS Sustain. Chem. Eng. **7**, 7277 (2019).
[128] H. Kivelä, M. Salomäki, P. Vainikka, E. Mäkilä, F. Poletti, S. Ruggeri, F. Terzi, and J. Lukkari, J. Phys. Chem. B **126**, 2324 (2022).
[129] W. Xu, E. I. Cooper, and C. A. Angell, J. Phys. Chem. B **107**, 6170 (2003).
[130] A. P. Abbott, R. C. Harris, and K. S. Ryder, J. Phys. Chem. B **111**, 4910 (2007).
[131] Y. Wang, W. Chen, Q. Zhao, G. Jin, Z. Xue, Y. Wang, and T. Mu, Phys. Chem. Chem. Phys. **22**, 25760 (2020).
[132] F. Stickel, E. W. Fischer, and R. Richert, J. Phys. Chem. **104**, 2043 (1996).
[133] P. Walden, Z. Physik Chem. **55**, 207 (1906).
[134] C. Schreiner, S. Zugmann, R. Hartl, and H. J. Gores, J. Chem. Eng. Data **55**, 1784 (2009).
[135] C. A. Angell, C. T. Imrie, and M. D. Ingram, Polym. Int. **47**, 9 (1998).
[136] R. Kind, O. Liechti, N. Korner, J. Hulliger, J. Dolinsek, and R. Blinc, Phys. Rev. B **45**, 7697 (1992).
[137] J. F. Douglas and D. Leporini, J. Non-Cryst. Solids **235**, 137 (1998).
[138] K. L. Ngai, J. Phys. Chem. B **110**, 26211 (2006).
[139] A. Dehaoui, B. Issenmann, and F. Caupin, Proc. Natl. Acad. Sci. **112**, 12020 (2015).
[140] C. M. Roland, S. Hensel-Bielowka, M. Paluch, and R. Casalini, Rep. Prog. Phys. **68**, 1405 (2005).
[141] G. P. Johari and O. Andersson, J. Chem. Phys. **125**, 094501 (2006).
[142] E. I. Cooper and C. A. Angell, Solid State Ionics, **18-19**, 570 (1986).
[143] Y. Hinz and R. Böhmer, J. Chem. Phys. **159**, (2023) https://doi.org/10.1063/5.0177377

# Supplementary Material
# for
# Translational and reorientational dynamics in carboxylic acid-based deep eutectic solvents

_______________________________________________________________________________

A. Schulz[1,a], K. Moch[2], Y. Hinz[2], P. Lunkenheimer[1], and R. Böhmer[2]

_______________________________________________________________________________

**AFFILIATIONS**

[1]Experimental Physics V, Center for Electronic Correlations and Magnetism, University of Augsburg, 86159 Augsburg, Germany
[2]Fakultät Physik, Technische Universität Dortmund, 44221 Dortmund, Germany

[a)]**Author to whom correspondence should be addressed:** arthur.schulz@physik.uni-augsburg.de

_______________________________________________________________________________

## 1. Time-dependent parameters obtained for maline

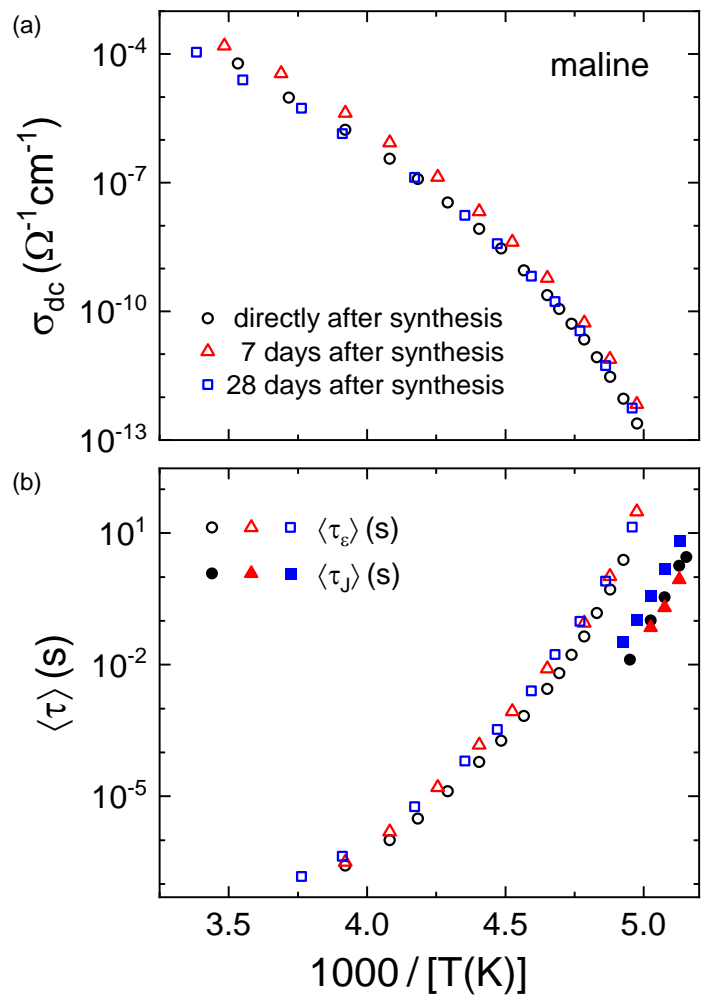

**FIG. S1.** Arrhenius representation of (a) the dc conductivity $\sigma_{dc}$ and (b) of the average dipolar and mechanical relaxation times $\langle\tau_\varepsilon\rangle$ and $\langle\tau_J\rangle$, respectively, as obtained from fits of the dielectric and rheological spectra of maline. The shown data were collected directly after synthesis (circles), 7 days after synthesis (triangles), and 28 days after synthesis (squares). Systematic changes cannot be observed in the obtained parameters across the investigated time frame.

## 2. Dielectric spectra for oxaline and phenylaceline

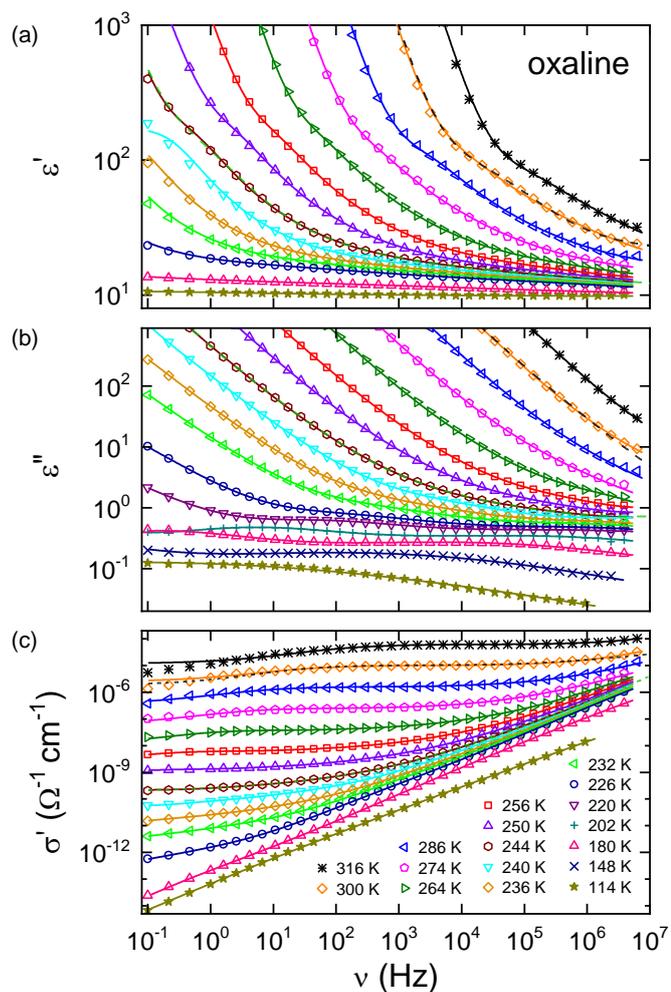

**FIG. S2.** This plot is analogous to Fig. 1, but for oxaline. The black and green dashed lines represent alternative fits of the 300 and 244 K spectra, respectively, which employ the Cole-Cole function to describe the $\alpha$ relaxation. They are of similar quality as the fits shown by the solid lines, that employ a Cole-Davidson function instead.

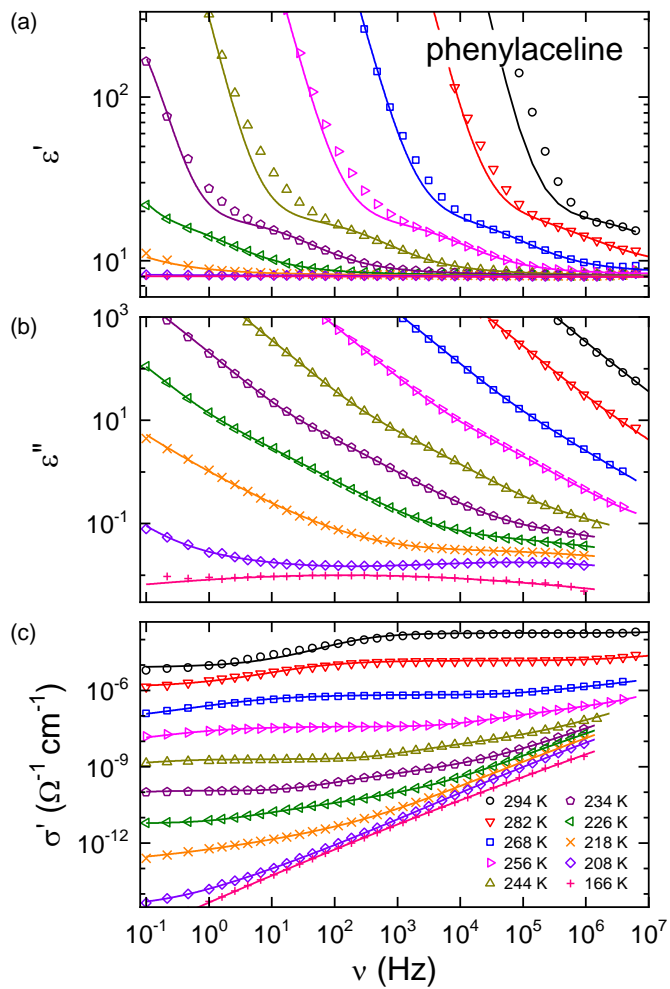

**FIG. S3.** This plot is analogous to Fig. 1, but for phenylaceline.

## 3. Rheological spectra for oxaline

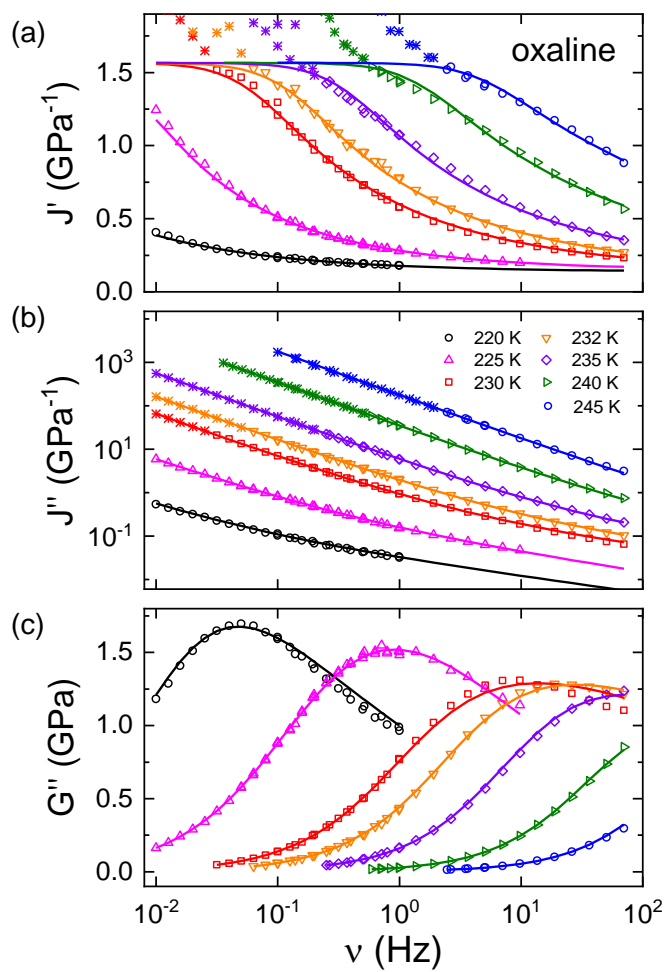

**FIG. S4.** This plot is analogous to Fig. 2, but for oxaline. The solid lines in (a) and (b) result from simultaneous fits of $J'$ and $J''$ on the basis of Eq. (2) with the parameters $J_\infty = 0.2 \pm 0.02$ GPa$^{-1}$, $\Delta J = 1.1 \pm 0.1$ GPa$^{-1}$, and $\beta_J = 0.30 \pm 0.05$.